\documentclass[journal=jacsat,manuscript=article]{achemso}
\usepackage[version=3]{mhchem}
\usepackage{graphicx}

\author{Kuldeep Kumar,\, P.Arun}
\affiliation[S.G.T.B. Khalsa College, University of Delhi]
{Mater. Sci. Res. Lab., S.G.T.B. Khalsa College, University of Delhi, 
Delhi 110007, INDIA}
\email{arunp92@physics.du.ac.in}
\author{Chhaya Ravi Kant}
\affiliation[Indira Gandhi Institute of Technology, Guru Gobind Singh 
Indraprastha University]
{\\Dept. of Appl. Sci., Indira Gandhi Institute of
Technology, G.G.S.I. University, Delhi 110 006}
\author{Bala Krishna Juluri}
\affiliation[Pacific Integrated Energy Inc.]
{\\Pacific Integrated Energy Inc., San Diego, California 92130, USA}

\title[\texttt{achemso} demonstration]
{Contribution of Halides in CsX (X=Cl, Br and I) Thin Films on the formation 
of SPR peaks}

\begin{document}
\begin{abstract}
This manuscript describes the evolution of grain structures in Cesium
Halide films which results in appearance of Surface Plasmon Resonance (SPR)
peaks in UV-visible absorption spectra. Thin films of Cesium Halide grown by 
thermal evaporation are polycrystalline in nature with large irregular 
tightly packed grains with sharp grain
boundaries. With time these grain boundaries recede to become smaller and
spherical in shape in order to minimize the free surface energy of the
system. The process is also assisted by the presence of point defects/color
centers in Cesium Halide films. These defects which appear due to the
absence of massive halide atoms from the lattice results in residual tensile
stress within the lattice which results in defect diffusion. The defects
migrate outwards towards the grain surface giving rise to Cesium metal
clusters. The lattice mismatch appearing between the lattices at the grain
boundary and the bulk giving rise to Core-shell structure further
contributes to grain division. Along with this volume diffusion, a surface
diffusion of Cesium takes place towards sites of faceted grain boundaries
resulting in a necking phenomenon, appearing like bridges between the daughter
Cesium Halide grains (grains appearing due to division of a single grain). 
Breaking away of the daughter Cesium Halide
grains results in nano-rods that contribute to the SPR peak seen.  

\end{abstract}
\newpage
\section{Introduction}

In our recent works we have reported about the optical properties of Cesium 
Halide thin films, namely, Cesium Chloride (CsCl), Cesium Bromide (CsBr) and 
Cesium Iodide (CsI). The optical properties stood out due to the singular
appearance of Surface Plasmon Resonance (SPR) peaks in the visible region.
While we confirmed that SPR peaks arise due to formation of
Cesium metal clusters, two basic questions arise from these observations, 
{\sl ``What mechanisms lead to the formation of cesium metal nano-clusters
especially the observed nano-rods?''} and {\sl ``Why does CsI behave 
differently from CsCl and CsBr?''} The direct experimental investigation into 
the formation of metal nano-clusters is not possible, however, we may 
logically speculate the sequence of events that led to the formation of 
metal nano-clusters.

\section{Results and discussion}

Thin films of cesium halide were fabricated by thermal evaporation in
vacuums better than ${\rm 10^{-5}}$~Torr. The films were deposited on 
microscopy glass slides maintained at room temperature. All the films were 
fabricated in identical conditions. Looking
at fig~1, an immediate observation is the striking similarity 
between morphology of CsCl, CsBr (not shown here) and CsI polycrystalline 
thin films. Large grains are tiled and tighly packed with sharp grain 
boundries. The grains however do not have regular shapes. These SEM 
micrographs were taken within few hours of sample fabrication. The 
morphologies gradually change with time as neighbouring grains receed to 
become smaller and spherical in appearance (fig~2).
%%%%%%%%%%%%%%%%%%%%%%%%%%%%%%%%%%%%%%%%%%%%%%%%%%%%%%%%%%%%%%%%%%%%
\begin{figure}[h]
\begin{center}   
\includegraphics[width=2.6in]{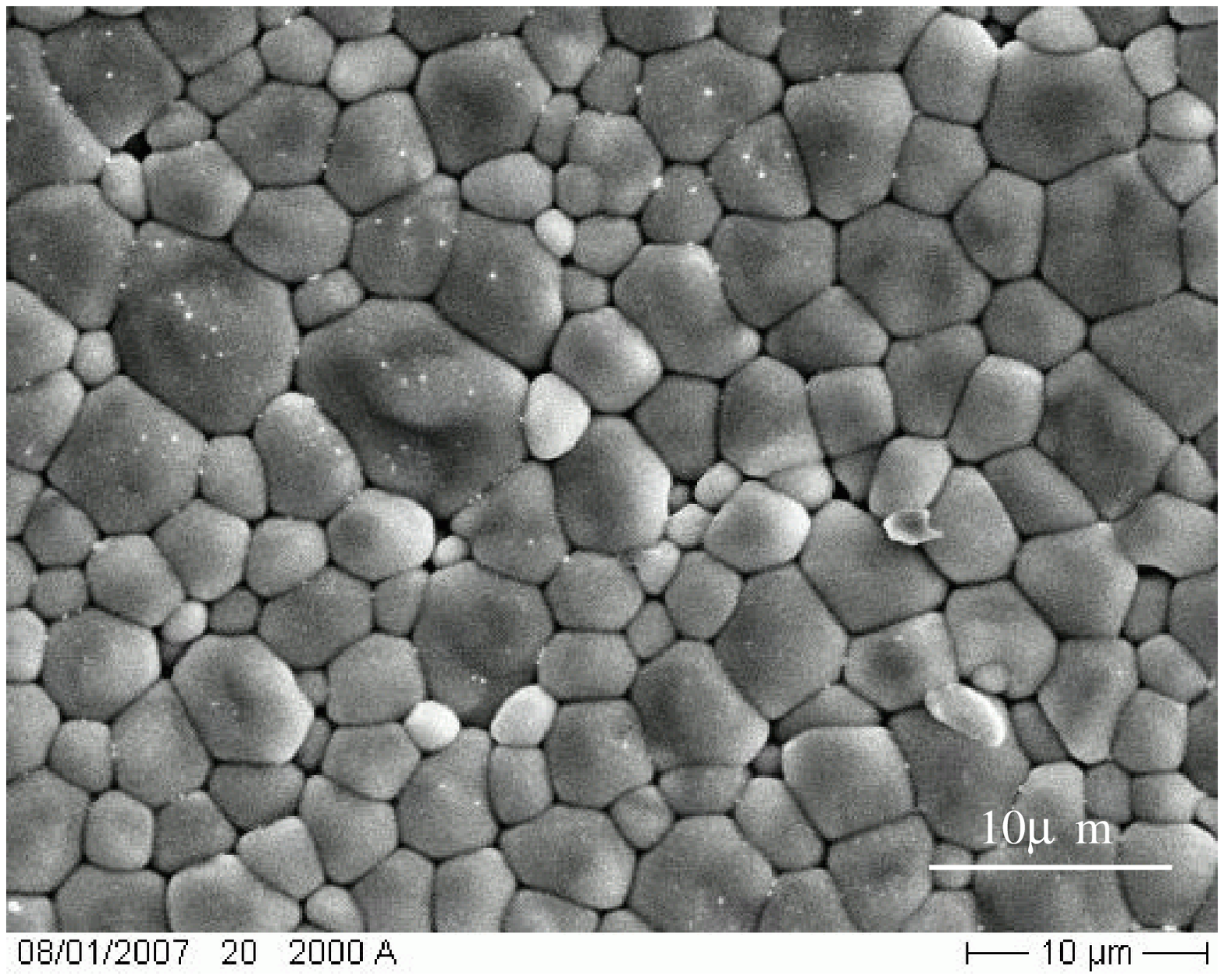}
\includegraphics[width=2.7in]{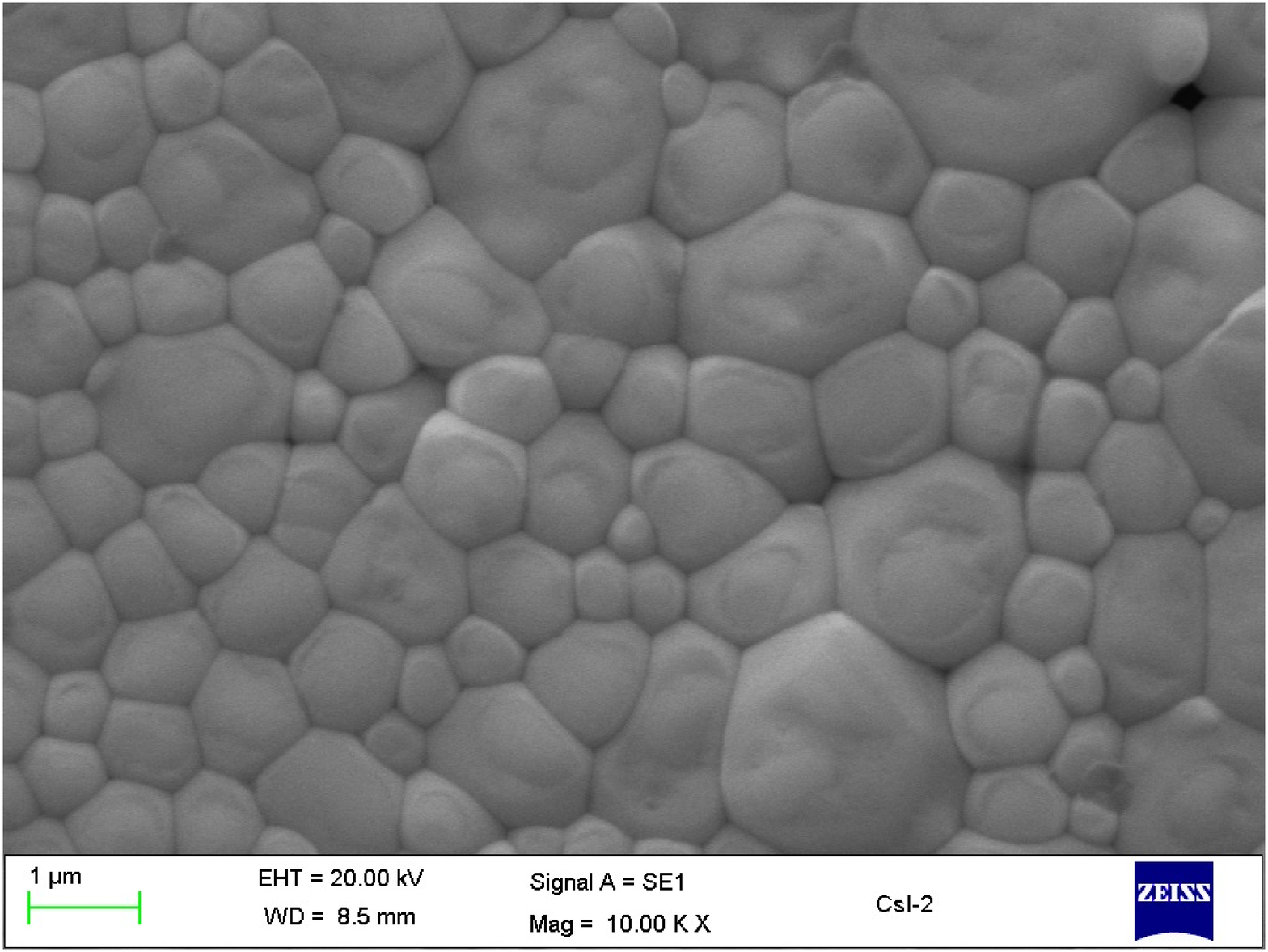}
\caption{\sl SEM images show the surface morphology
(polycrystalline nature) of thin films of Cesium Chloride and Cesium Iodide
respectively.}
\end{center}  
\label{tiles}  
\end{figure}  
%%%%%%%%%%%%%%%%%%%%%%%%%%%%%%%%%%%%%%%%%%%%%%%%%%%%%%%%%%%%%%%%%%%%%%%%%%%%%%
%%%%%%%%%%%%%%%%%%%%%%%%%%%%%%%%%%%%%%%%%%%%%%%%%%%%%%%%%%%%%%%%%%%%
\begin{figure}[h!!!]
\begin{center}
\includegraphics[width=2.45in]{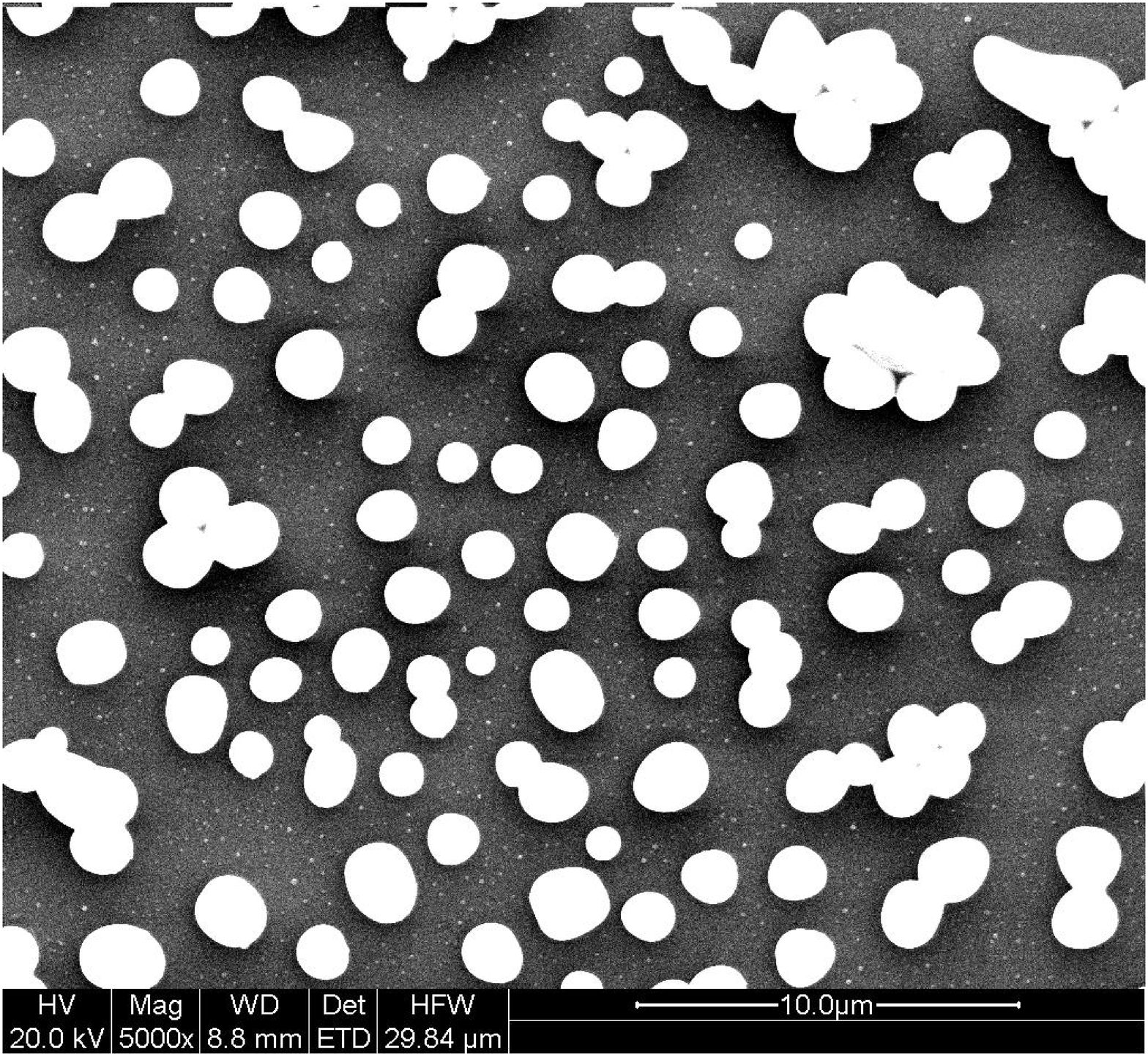}
\hfil
\includegraphics[width=3.0in]{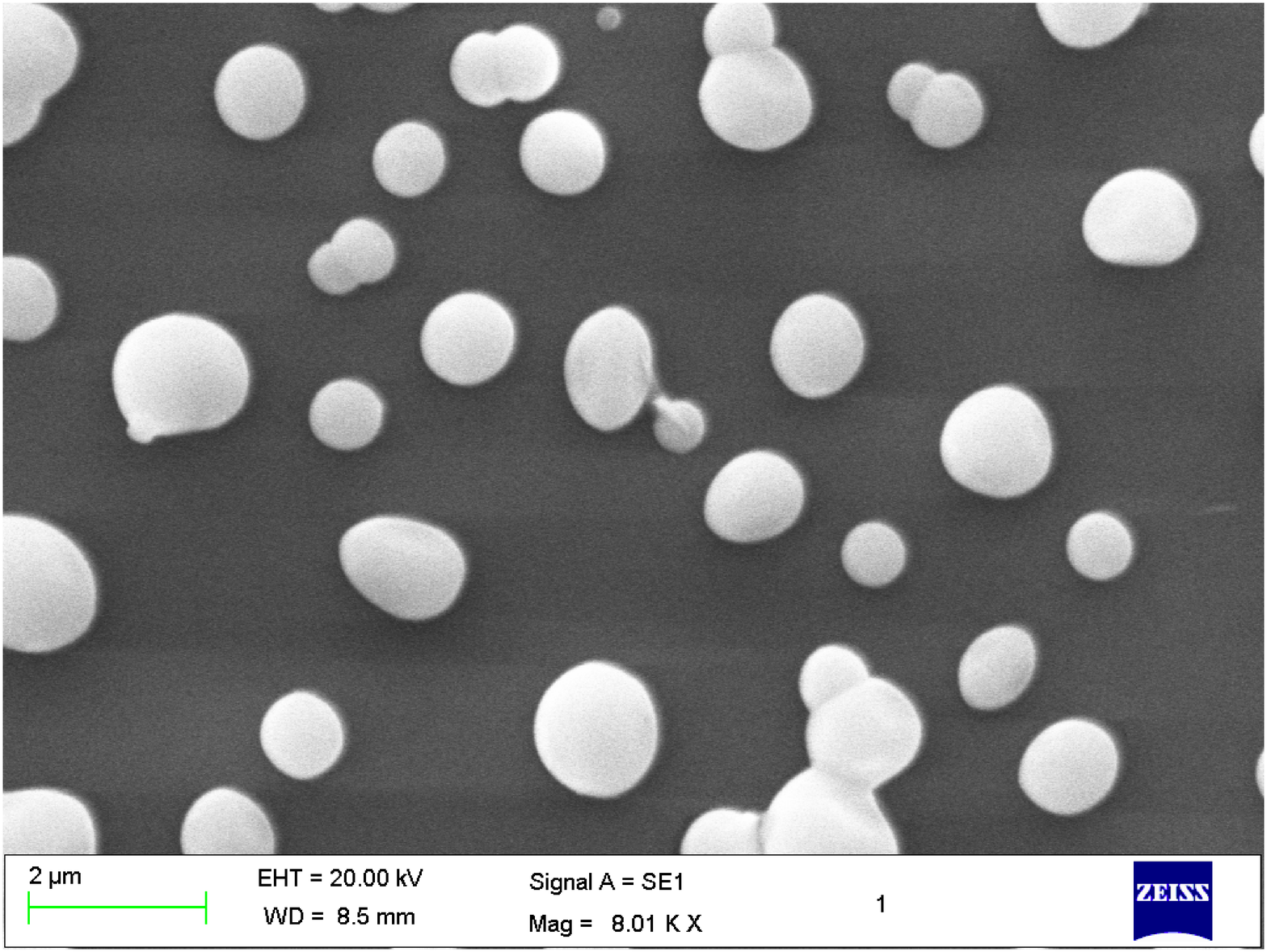}
\caption{\sl SEM morphologies gradually change with time as neighbouring
grains receed the become smaller and spherical in appearance (Images are of
CsCl and CsI of samples shown in fig~1 after ageing.}
\end{center}
\label{fig2}
\end{figure}
%%%%%%%%%%%%%%%%%%%%%%%%%%%%%%%%%%%%%%%%%%%%%%%%%%%%%%%%%%%%%%%%%%%%%%%%%%%%%%

\subsection{Grain Boundary Grooving: Surface Diffusion}

Nature has a natural tendency to force grains to take up a spherical shape.
This is nature's way of minimizing the free energy of the system 
(G)~\cite{xin}. The equation used to model this behavior of nature is given 
as
\begin{eqnarray}
G=\gamma_sA_s +\gamma_bA_b\label{groove}
\end{eqnarray}
where ${\rm \gamma_s}$ and ${\rm \gamma_b}$ are the surface energy per unit
area and grain boundary energy per unit area respectively. ${\rm A_s}$ and
${\rm A_b}$ being the grain's surface area and grain boundary area 
respectively. Grain boundary area is the area of contact between two grains.
For a spherical grain of a given volume, the system's free 
energy is the least due to a large decrease in grain boundary area,   
${\rm A_b}$ at the cost of a relatively small increase in surface area
(${\rm A_s}$).
%%%%%%%%%%%%%%%%%%%%%%%%%%%%%%%%%%%%%%%%%%%%%%%%%%%%%%%%%%%%%%%%%%%%
\begin{figure}[h!!!]
\begin{center}
\includegraphics[width=3in]{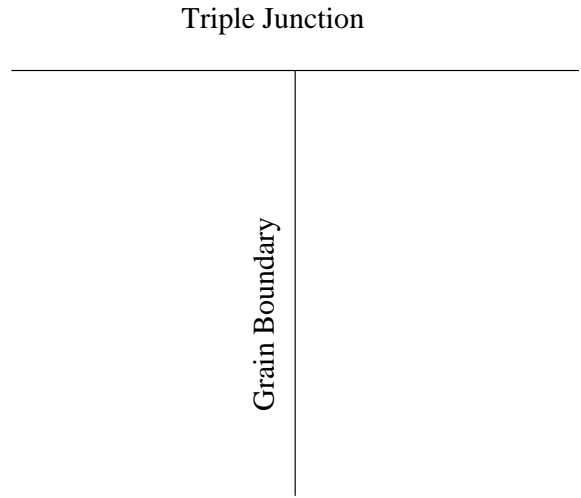}
\caption{\sl Schematic representation of the triple junction, the point
where grain boundary meets the surface. }
\end{center}
\label{triplejun}
\end{figure}
%%%%%%%%%%%%%%%%%%%%%%%%%%%%%%%%%%%%%%%%%%%%%%%%%%%%%%%%%%%%%%%%%%%%%%
The grain shaping or its change to spherical shape is due to grain boundary
grooving. ``Grooving'',\cite{mull,her,tsoga} as the name suggests, is the
phenomenon
by which gaps appear between the grains of polycrystalline samples. The   
fissures develop from the film surface towards the substrate. Theoretical 
models show the grooving occurs due to atoms moving along the surface away
from the ``triple intersection point'' by surface diffusion~\cite{sun}  
(fig~3). Fig~1, clearly shows due to the near hexagonal tiling of the
grains, triple intersection points are formed even within the film. Fig~4
shows pictorial representation of fig~1 (initial situation is shown in
inset) followed by how these triple intersection points evolve, pushing into
the grain and thus giving a sphereical shape.

%%%%%%%%%%%%%%%%%%%%%%%%%%%%%%%%%%%%%%%%%%%%%%%%%%%%%%%%%%%%%%%%%%%%
\begin{figure}[h!!!]
\begin{center}
\includegraphics[width=3in]{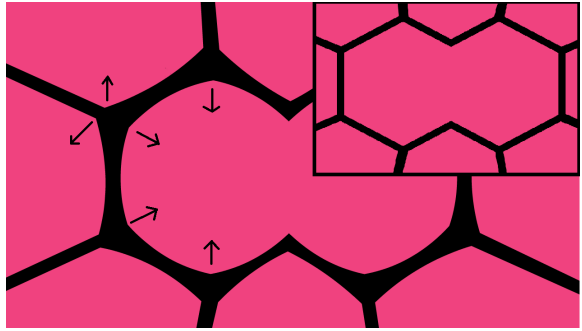}
\caption{\sl Pictorial representation of how triple intersection points
evolve, pushing grains to have spherical shapes. Arrows are for indication
of the direction in which these triple points evolve.}
\end{center}
\label{t1}
\end{figure}
%%%%%%%%%%%%%%%%%%%%%%%%%%%%%%%%%%%%%%%%%%%%%%%%%%%%%%%%%%%%%%%%%%%%%%

In the above paragraph, we have assumed ${\rm \gamma_b}$ to be a constant. This 
is far from the truth since the grain boundary energy depends on the lattice 
defects caused by the mismatch between the lattice at the grain boundary and 
its bulk~\cite{frank,bass,maks,stra,maks1,gari}. This mismatch manifests as a 
shell region whose thickness is proportional to the grain size. Smaller grains 
have 
thinner shells with less lattice mismatch and hence smaller ${\rm \gamma_b}$. 
Thus nature breaks the grains into smaller spherical grains with time, thus 
reducing the system's free energy by lowering 
${\rm \gamma_b}$.\cite{bouville2} Fig~5 highlights the formation of shell
around the grain via volume diffusion of defect towards the surface. We shall 
be discussing volume diffusion further, however at this point it is important
to appreciate that if there is a lattice mismatch between the shell and core
then ${\rm \gamma_b}$ takes values to assists grooving of the grains.

%%%%%%%%%%%%%%%%%%%%%%%%%%%%%%%%%%%%%%%%%%%%%%%%%%%%%%%%%%%%%%%%%%%%%%%%%%
\begin{table}[h]
\begin{center}
\caption{Table for Lattice mismatch in core-shell structure}
\vskip 0.5cm
\begin{tabular}{lllll}
%\hline\\ 
\hline
Halides & Core structure & Lattice constant & Shell structure & Lattice
constants \\ \hline 
Chlorine & Cubic & 4.12 & Cubic & 5.838  \\
Bromine & Cubic & 4.29 & Cubic & 5.984  \\ 
Iodine & Cubic & 4.568 & Tetragonal & 3.3645,3.4645,12.552   \\ \hline

\end{tabular} 
\label{tab1}
\end{center}
\end{table}  
%%%%%%%%%%%%%%%%%%%%%%%%%%%%%%%%%%%%%%%%%%%%%%%%%%%%%%%%%%%%%%%%%%%%%%%%%%

%%%%%%%%%%%%%%%%%%%%%%%%%%%%%%%%%%%%%%%%%%%%%%%%%%%%%%%%%%%%%%%%%%%%
\begin{figure}[h!!!]
\begin{center}
\includegraphics[width=3in]{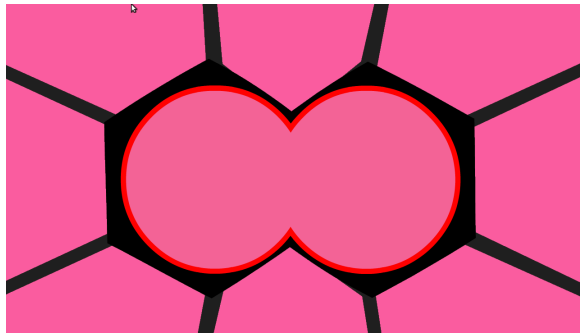}
\caption{\sl Pictorial representation of shell formation.}
\end{center}
\label{t3}
\end{figure}
%%%%%%%%%%%%%%%%%%%%%%%%%%%%%%%%%%%%%%%%%%%%%%%%%%%%%%%%%%%%%%%%%%%%%%

As described above, the lattice mismatch of the two regions contributes to 
${\rm \gamma_b}$. In our case, this lattice mismatch is due to a core-shell formation
taking place with Cesium shell around a core of Cesium halide. The
existence of free metallic Cesium were shown by the just resolvable XRD
peaks.\cite{kapil1, kapil2} 
The XRD results also allowed us to determine the 
lattice structure and size for the two regions. Table~1 lists the 
results for easy viewing. Larger mismatch in Cesium and Cesium Halide
structures imply greater ${\rm \gamma_b}$. Since, nature tries to bring down 
the system's free energy by reducing ${\rm \gamma_b}$ via grain divison, the 
shells become thinner, thus reducing the region of lattice mismatch.
Since the cesium shell in CsI takes a tetragonal structure, which is 
very different from the cubic structured core, it would necssarily indicate a 
faster grain division/breakage in CsI as compared to CsCl and CsBr.

\subsection{Core-Shell Formation: Volume Diffusion}

As stated above, a thin layer (shell) of Cesium is formed around the Cesium 
halide core in our samples. The question is, {\sl ``how do the halide atoms 
disappear from the surface or possibily how does cesium reach the surface 
with no halide atom?''} While iodine might have tendency to sublimate, leaving
behind free Cesium, the remaining halides do not sublimate. 
Hence, the question of importance is {\sl ``how does cesium clusters form at 
the grain surface?"} Cesium Halides crystals/ films readily form color centers. 
These are point defects due to the absence of the massive halides atoms from
the lattice which results in residual tensile stress within the lattice. The 
neighbouring lattices inturn experinces compressive stress. The resulting 
stresses graidients in turn result in diffusion of defects~\cite{gari}. 

Consider the diffusion process of a vacancy. From its intial position to
final poistion, the vacancy moves through intermediate steps which is 
essentially marked by the vacancy choosing the highest free energy state and 
minimum energy diffusion path. The difference in the free energy between the 
intermediate steps is called migration energy ($E_{mig}$). The process of 
defect diffusion depends on (i) formation of defect ($E_{form}$) and 
(ii) its migration. Hence we can talk of an activation energy for defect
diffusion which would be the sum of these energies, i.e.
\begin{eqnarray}
A=E_{form}+ E_{mig}\nonumber
\end{eqnarray}
The diffusivity in terms of activation energy is given by~\cite{zangwill}
\begin{eqnarray}
D=\frac{f\nu d^2}{6}exp\left(\frac{-A}{KT}\right)\label{gana}
\end{eqnarray}
where, `f' is a correlation factor (propostionality constant), $\nu$ the 
attempt frequency, `d' the hop distance. The vacancies due to the stress 
gradient migrate outward towards the grain surface, leading to an 
accumulation of point defects at the surface. These point defect then would 
combine to give Cs metal clusters.

As a starting point of discussion, in eqn(\ref{gana}) let us assume the    
activation energy of all the three cesium halides to be comparable. We would
find the diffusion coefficient to be proportional to
\begin{eqnarray}
D_{halide} \propto vd^2\nonumber
\end{eqnarray}
The hoping frequency, `$v$', or the rate at which halide atoms would migrate  
to give vacancy's diffusion, would depand on the halide atom's inertia or 
mass. The heavier the atom would have lower frequency of hoping. Hence,  
`$v$' would be inversely proportional to the halide atoms mass.

Also, `$d$', 
the hoping distance would be directly proportional to the lattice constant.    
A plot between the lattice constant and halide atomic mass shows a linear  
relationship (fig~6). This is expected since, heavier atoms mean larger  
radius and in turn larger lattice dimensions. The atom's migration for 
filling vacancy (hence give illusion of vacancy migration) would be 
discouraged with increasing distance. Hence, `$d$' should show an inverse 
proportionality with mass, however, it should be noted here that the lattice 
size increases more rapidly than the radius of the halide atom as we move 
from  CsCl to CsI. This would mean that it would be easier for the iodine atom to
move from one unit cell to another as compared to chlorine atom in its   
lattice. Hence, taking these into account, `d' is independent of halide  
atom's mass (${\rm d \propto {1 \over m}\times m =constant}$). Hence, the
diffusion coefficient (eqn~2) is inversely proportional to the 
concerned halide atom's mass
%%%%%%%%%%%%%%%%%%%%%%%%%%%%%%%%%%%%%%%%%%%%%%%%%%%%%%%%%%%%%%%%%%%%%%%%%
\begin{figure}[h]
\begin{center}   
\includegraphics[width=2.0in, angle=-90]{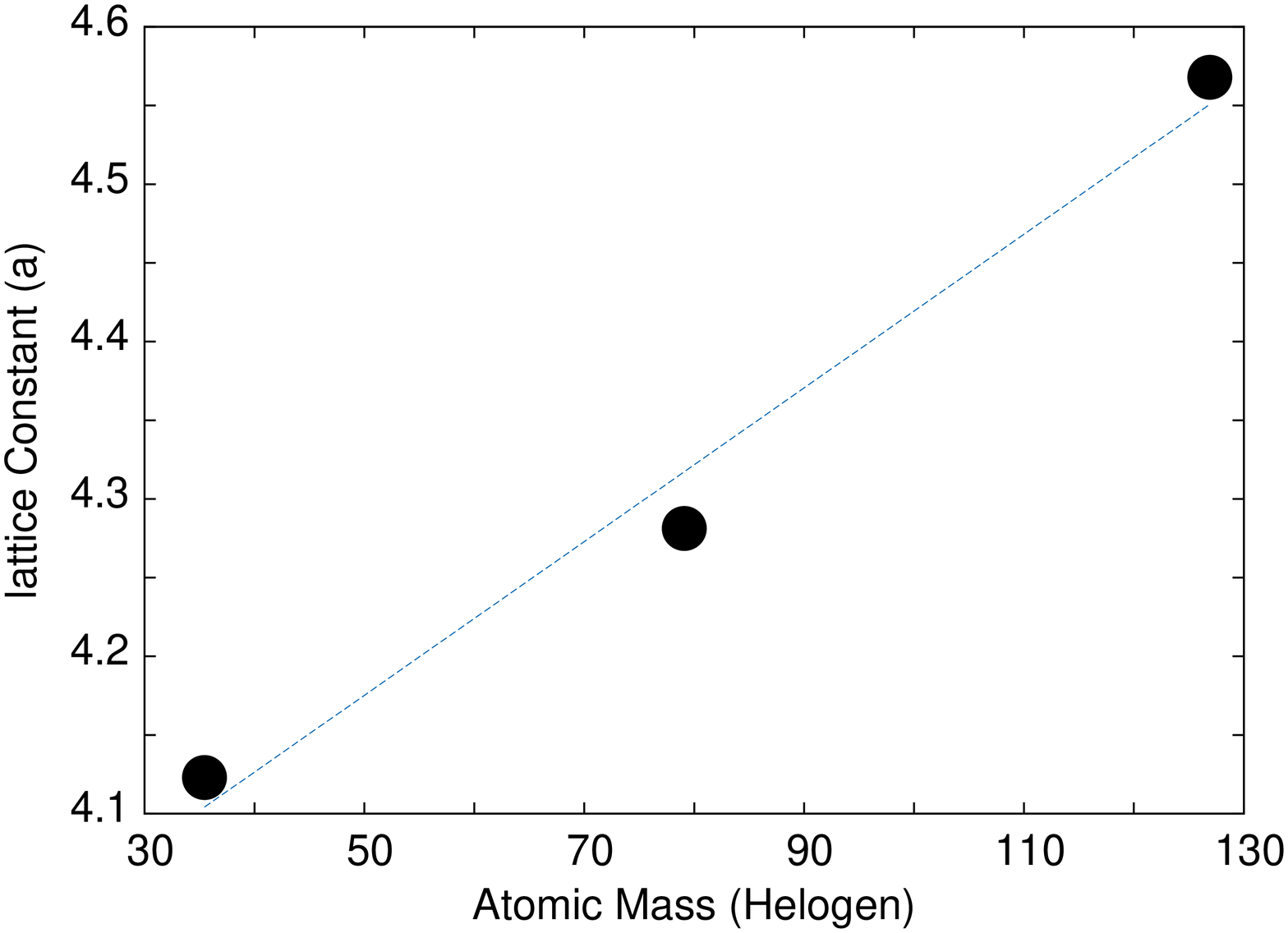}
\hfil
\includegraphics[width=2.0in, angle=-90]{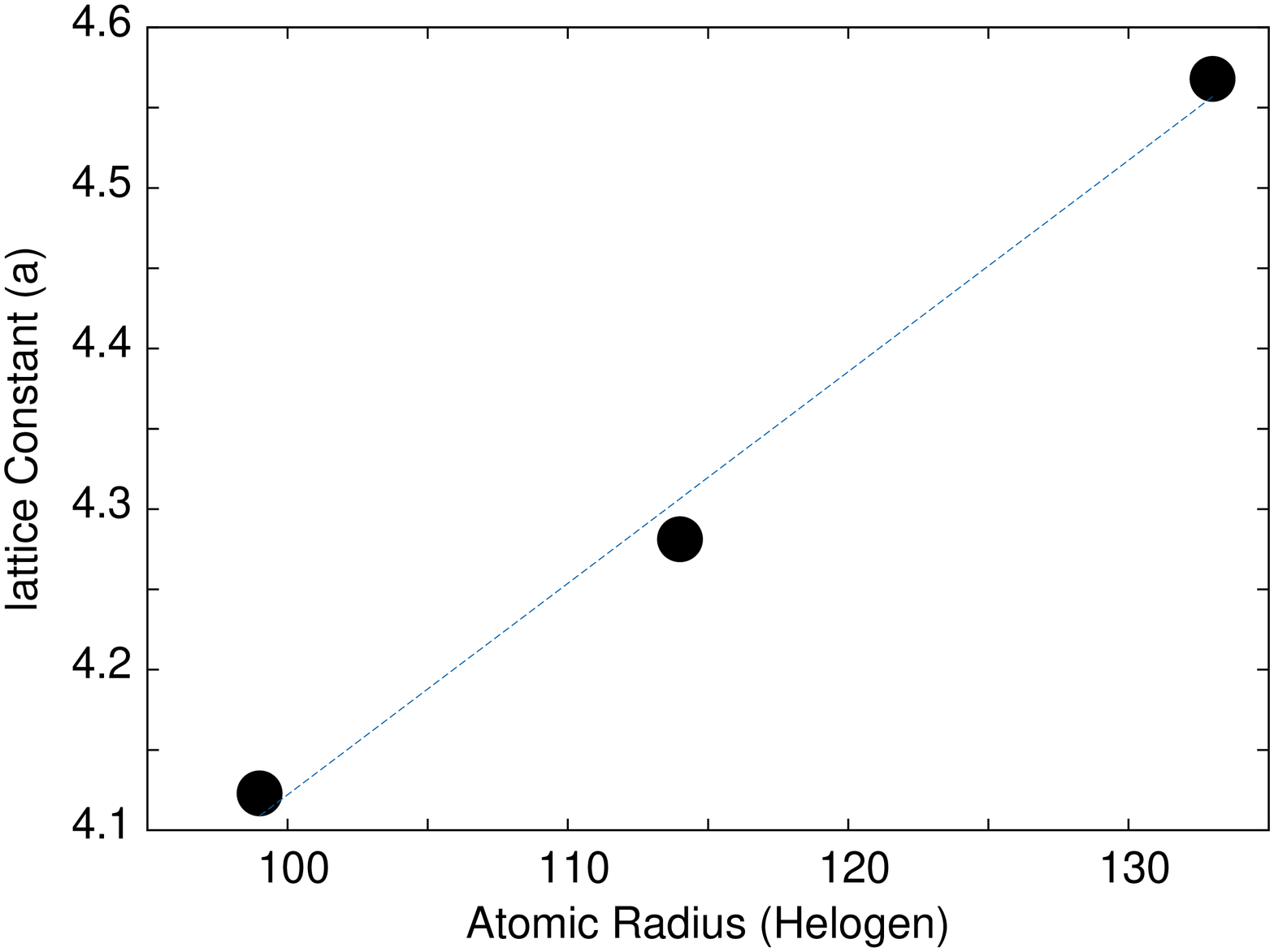}
\caption{\sl Variation between halogen atomic mass and lattice constant of
Cesium halides}
\end{center}   
\label{fig11}  
\end{figure}   
%%%%%%%%%%%%%%%%%%%%%%%%%%%%%%%%%%%%%%%%%

\begin{eqnarray}
D_{halide} &\propto & \left(1 \over m \right) \left(1 \over m^2
\right)m^2\nonumber\\
&\propto & 1 \over m  \label{diff}
\end{eqnarray}
From the listings in Table~2 it is clear $D_{Cl}\,>\,D_{Br}\,>\,D_I$.
This would imply that the cesium shell formation would be fastest in CsCl,
followed by CsBr and finally CsI.

\subsection{Formation of Nano-rods: Surface Diffusion/ Necking}
As the grooving is taking place and the large grains are separating into two 
daughter grains, surface diffusion of cesium metal also takes place with the 
volume diffusion of defects. This surface diffusion takes place towards sites 
of
faceteds on the grain boundary. This assists in accumulation of Cesium that
pushes the daughter grains apart and formation of a bridge of Cesium between
the two grains (see fig~7). As the distances between the two grains
increases this bridge is elongated.  
%%%%%%%%%%%%%%%%%%%%%%%%%%%%%%%%%%%%%%%%%%%%%%%%%%%%%%%%%%%%%%%%%%%%%%%%%
\begin{figure}[h]
\begin{center}   
\includegraphics[width=3.0in]{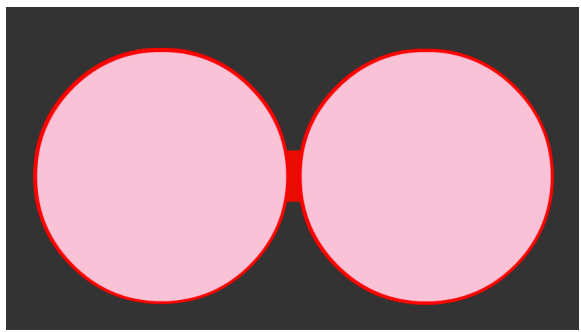}
\caption{\sl Pictorial representation of ``neck" formation.}
\end{center}   
\label{t5}  
\end{figure}   
%%%%%%%%%%%%%%%%%%%%%%%%%%%%%%%%%%%%%%%%%

Transmission Electron micrograph of fig~8, shows the bridges/ nano-rod 
formations between two evolving ``daughter grains'' of Cesium Bromide. These 
bridges are called ``necks", marking the region of 
constricted areas joining two grains. As stated, necking is usually
formed at sites of faceted grain boundaries \cite{faceted1, faceted2}. With
a core-shell structure in all three halides, formation in necking is
expected. The necks are marked by difference in
curvature. While the grain surfaces are convex in nature, necks have concave
surface~\cite{brown1may}. Matter flows from the grain surface to region of
maximum variation of curvature, i.e. towards the neck. That is, the neck 
growth and elongation results due to surface diffusion~\cite{julien, maeno}.
Surface diffusion is different from volume diffusion (defects migrating
towards the surface). Since our grains have Cesium shells,
surface diffusion essentially ends up carrying Cesium towards the neck.
Hence, the bridges formed are made up of Cesium which on breaking free from
the daughter-grains
gives the nano-rods which contribute to the Surface Plasmon Resonance seen in
the UV-visible absorption pattern. 
%%%%%%%%%%%%%%%%%%%%%%%%%%%%%%%%%%%%%%%%%%%%%%%%%%%%%%%%%%%%%%%%%%%%%%%%%
\begin{figure}[t!!]
\begin{center}
\includegraphics[width=4.25in, angle=0]{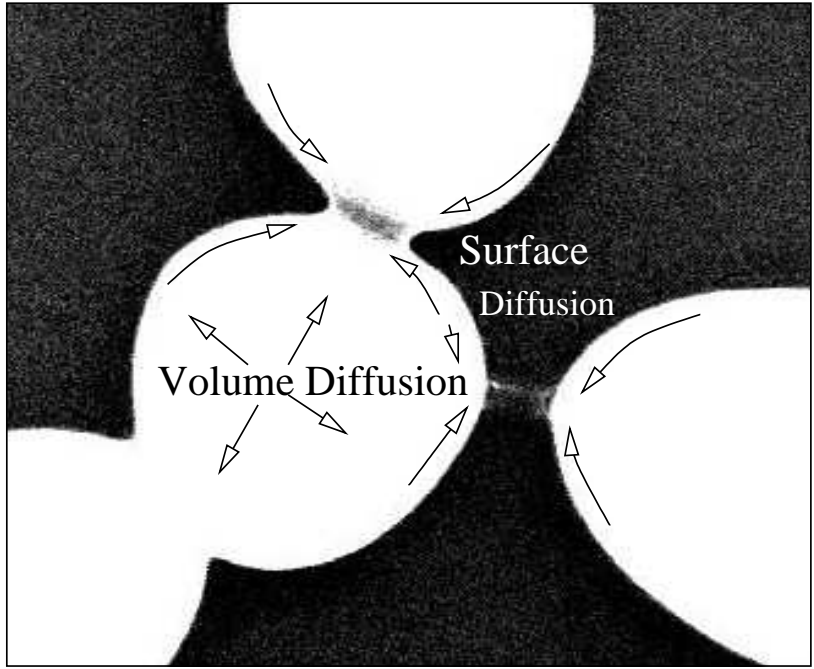}
\caption{\sl The two diffusions, volume and surface diffusion contribute to
the forming of spherical grains with Cesium shell in Cesium halides.}
\end{center}
\label{thermal}
\end{figure}   
%%%%%%%%%%%%%%%%%%%%%%%%%%%%%%%%%%%%%%

Another important observation we reported was the ageing of our samples.
That is, the optical properties of the films were found to change with time.
The absorption peaks not only red-shifted but also their
instensities diminished with time. Also, we observed that the rate of ageing
was sensitive to the ambient atmosphere in which the samples were
maintained, with ageing being more rapid in samples kept out of the
dessicator. Our Gan's model calculations explained the red-shift due to the
nanorods going smaller with lower aspect ratio. We believe the Cesium
nanorods and free Cesium on the grain surfaces are eroded by the reaction
\begin{eqnarray}
2Cs+2H_2O=2CsOH+H_2\nonumber
\end{eqnarray}

\begin{table}
  \caption{Table for various useful parameters}
  \label{tbl:example}
  \begin{tabular}{lllllll}
    \hline
	Halides & At. No. & Lattice Size & Coval. dist. & 
	At. Mass & Radius & M. Points \\ 
	\hline 
	Chlorine & 17& 4.12 & 198 & 35.457 & 99 &645 \\
	Bromine & 35 & 4.29 & 227 &  79.904 & 114 &636 \\
	Iodine & 53 & 4.56 & 272 & 126.9 & 133 & 621 \\ 
	\hline
  \end{tabular}
\end{table}

K. Arima et. al.~\cite{arima} have observed that this reaction preceeds
slowly for atomospheric relative humidity (RH) less than 30\% and increases
with RH. Also, there is a near exponential increase in the rate of reaction
for RH above 80\%. K. Arima et. al.~\cite{arima} 
works would explain difference in ageing observed by us in the samples
maintained in the dessicator and those left in open. Also, Antonelli et. 
al.~\cite{anton} have shown that vacancy generation increases
under the high hydrostatic pressure (compressive). Hence, metal Cs formation
is encouraged in dessicator with slower $Cs+H_2O$ reaction rate. While this
reaction involves only Cesium, we do observe a difference in ageing rate
depending on which halide is in bonding with Cesium. This is due to fact
that ageing depends on (i) what rate Cesium clusters are formed at the
surface (volume diffusion which is inversely dependent on halide mass) and
(ii) thicker shell leading to faster grooving (surface diffusion due to
lattice mismatch).

Comparing the ageing rate of 600~nm thin films of CsCl, CsBr and CsI, we
found the rate of ageing in the following order, $CsCl\,>\,CsBr$ and no
ageing in CsI. The
fastest ageing was seen in CsCl since color centers moved towards the
surface faster with volume diffusion rate being inversely proportional to
the halide atom's mass. With lattice mismatch being low, grooving was slower
and formation of nano-rods retarded. While in the case of CsI, formation of
Cesium shell would be slower, but grooving faster. The thinner Cesium shells
would not enable formation of large nanorods, giving impression of no ageing
in CsI.

\section{Conclusion}

To summerize, thermal evaporated films of the three halides result in 
polycrystalline films with large grains and sharp grain boundaries. The
grain boundaries become sites of grooving with mass of material being pulled 
away from the grain boundary to give spherical grains. Along with this, 
vacancies in the form of color centers in the
cesium halide lattices move towards the grain surface. The rate at which
this volume diffusion occurs depends on halide atom's mass (its faster in CsCl 
than CsBr or CsI). At the surface, these color centers accumulate, forming 
metal clusters and giving rise to a core-shell structure. Depending on the 
type of lattice mismatch in the core-shell
structure two alternative paths of sequencing occurs. While the screw
dislocation leads to necking, the mismatch in lattice size leads to an
accelerated grooving. Grooving is faster in CsI than CsCl and CsBr. Accelerated
grooving leads to smaller grains. Smaller grains encourage reaction with 
water vapor, thus reducing
the shell size. Making it difficult to sustain the necking. The necks break
off faster. Ageing observed in our samples is essentially
due to the nanorods of Cesium and in decreasing aspect ratio. The decreasing
aspect ratio, with time and inturn successive grain division would be due to
the smaller grain size. Smaller grain size would inturn imply thinner shells
and lesser amount of Cesium. This in turn would not allow for growth of
longer nanorods. Since CsI have a large mismatch in their
core-shell lattices, grooving is faster and shell is thinner compared to
CsCl and CsBr. Hence the nano-rods formed are of smaller length. This makes it   
difficult to obtain SPR peaks shifting throughout the full visible range
with ageing in CsI.

\acknowledgement

The authors would like to express their sincere gratitude
to Department of Science and Technology (DST) India (SR/NM.NS-28/2010) and
University Grants Commission (UGC, Delhi) (F.No. 39-531/2010 SR) for
the financial assistance given for carrying out this work.

%% \suppinfo macro.
%%%%%%%%%%%%%%%%%%%%%%%%%%%%%%%%%%%%%%%%%%%%%%%%%%%%%%%%%%%%%%%%%%%%%

\subsection{References}

%\bibliography{achemso}

\begin{thebibliography}{99}

\bibitem{mull} W.W. Mullins, J. Appl. Phys. {\bf 28} (1957) 333.
\bibitem{her} C. Herring, The Physics of Powder Metallurgy (McGraw-Hill,
New York, 1951), 143.
\bibitem{tsoga} A. Tsoga and P. Nikolopoulos, J. Am. Soc. {\bf 77} (1994) 954.
\bibitem{sun} B. Sun and Z. Suo, Acta Materialia, {\bf 45} (1997) 4953.
\bibitem{xin} T. Xin and H. Wong, Acta Mater., {\bf 51} (2003) 2305.
\bibitem{frank} F.C. Frank and J. H. Van der Merwe, {\bf 198} (1949) 205.
\bibitem{bass} L.C. Bassman, ``Modeling of Stress-Mediated Self-Diffusion in Polycrystalline Solids''. Ph.D. dissertation, Stanford University,  (1999).
\bibitem{maks} E.L. Maksimova, E.I. Rabkin, L.S. Shvindlerman, and B.B.
Straumal, Acta Metall. {\bf 37}, (1989) 1995.
\bibitem{stra} B.B. Straumal and L.S. Shvindlerman, Acta Metall. {\bf 33}, 
(1985) 1735.
\bibitem{maks1} E.L. Maksimova, L.S. Shvindlerman, and B.B. Straumal, Acta
Metall. {\bf 37}, (1989) 2855.
\bibitem{gari} K. Garikipati, L. Bassman and M. D. Deal, J. Mech. Phys. Sol., 
{\bf 49}, (2001) 1209.
\bibitem{bouville2} Mathieu Bouville, Shenyang Hu, Long-Qing Cheu, Dongzhi
Chi and David J Srolovitz, Modelling Simulation Mater. Sci. Eng., {\bf 14}
333 (2006).
\bibitem{kapil1} Kuldeep Kumar, P. Arun, Chhaya Ravi Kant and Vincent
Mathew, A. Phys. A., {\bf 99} 305 (2010).
\bibitem{kapil2} Kuldeep Kumar, P.Arun, Chhaya Ravi Kant and Bala Krishna
Juluri, A. Phys. Lett., {\bf 100} 243106 (2012).
\bibitem{faceted1} A. Ramasubramaniam and V.B. Shenoy, Acta Materialia, {\bf
53} 2943 (2005).
\bibitem{faceted2} Donghong Min and Harris Wong, J. Appl. Phys., {\bf 100},
053523 (2006).
\bibitem{brown1may} R.L. Brown and M.Q. Edens, J Glaciology, {\bf 37}, 203
(1991).
\bibitem{julien} Julien Bruchon, Daniel Pino-Munoz, Francois Valdivieso and
Sylvain Drapier, J Am Ceram Soc, {\bf 95} 2398 (2012).
\bibitem{maeno} N. Maeno and T. Ebinuma, J. Phys. Chem., {\bf 87},
4103 (1983).
\bibitem{arima} K. Arima, P. Ziang, X. Deng, H. Bluhm and M. Salmeron, J of
Phys Chem C, {\bf 114}, 14900 (2010).
%\bibitem{zangwill} A. Zangwill, {\sl ``Physics of Surface''} (Cambridge
%University Press, UK 1988).
\bibitem{anton} A. Antonelli and J. Bernholc, Pressure and Strain Effects on Diffusion, MRS Proceedings, {\bf 163}, (1989) 523.
 
\end{thebibliography}

\end{document}